\documentclass[iop,apj,numberedappendix,onecolumn]{emulateapj}
\usepackage{natbib}
\usepackage{graphicx}
\usepackage{amsmath,amsfonts}
\usepackage{booktabs}
\usepackage{hyperref}
\usepackage{color} 

\hypersetup{colorlinks=false,pdfborder={0 0 0}}

\slugcomment{Accepted for publication in ApJ, June 18, 2016}

\setcitestyle{notesep={; }}

\def\sgra{Sgr~A$^{\ast}$}

\def\lsim{\mathrel{\raise.3ex\hbox{$<$\kern-.75em\lower1ex\hbox{$\sim$}}}}
\def\gsim{\mathrel{\raise.3ex\hbox{$>$\kern-.75em\lower1ex\hbox{$\sim$}}}}
\def\gtwid{\mathrel{\raise.3ex\hbox{$>$\kern-.75em\lower1ex\hbox{$\sim$}}}}
\def\proptwid{\mathrel{\raise.3ex\hbox{$\propto$\kern-.75em\lower1ex\hbox{$\sim$}}}}

\begin{document}

\title{ The Optics of Refractive Substructure }
\shorttitle{The Optics of Refractive Substructure}

\author{Michael D.~Johnson and Ramesh Narayan}
\shortauthors{Johnson \& Narayan}
\affil{Harvard-Smithsonian Center for Astrophysics, 60 Garden Street, Cambridge, MA 02138, USA}

\email{mjohnson@cfa.harvard.edu} 


\begin{abstract}
Newly recognized effects of refractive scattering in the ionized interstellar medium have broad implications for very long baseline interferometry (VLBI) at extreme angular resolutions. Building upon work by \citet{Blandford_Narayan_1985}, we present a simplified, geometrical optics framework, which enables rapid, semi-analytic estimates of refractive scattering effects. We show that these estimates exactly reproduce previous results based on a more rigorous statistical formulation. We then derive new expressions for the scattering-induced fluctuations of VLBI observables such as closure phase, and we demonstrate how to calculate the fluctuations for arbitrary quantities of interest using a Monte Carlo technique. 
\end{abstract}

\section{Introduction}

Very long baseline interferometry (VLBI) now provides angular resolution of tens of microarcseconds, both at millimeter wavelengths with the Event Horizon Telescope \citep[EHT;][]{Doeleman_2009} and at centimeter wavelengths with {\it RadioAstron} \citep{Kardashev_2013}. As these and future interferometers push to ever higher angular resolution, they must contend with a fundamental limitation: radio-wave scattering in the inhomogeneous, ionized interstellar medium (ISM). It has generally been assumed that the effects of scattering on resolved images of AGN are simply to ``blur'' the images with an approximately Gaussian kernel that grows roughly with the squared observing wavelength. However, within individual observing epochs scattering has the opposite effect, introducing substructure into the scattered image \citep{NarayanGoodman89,GoodmanNarayan89,Johnson_Gwinn_2015}. Unlike the time-averaged blurring of scattering, the substructure becomes stronger at shorter wavelengths and does not correspond to a convolution of the unscattered image. The scattering substructure predominantly affects visibilities on long baselines and can significantly influence high-resolution VLBI data. 

Scattering is especially important for studies of the Galactic Center supermassive black hole, Sagittarius~A$^\ast$ (\sgra), with the EHT and for studies of extremely high brightness temperatures (${\gg}\,10^{12}~{\rm K}$) in active galactic nuclei (AGNs) with {\it RadioAstron}. For \sgra\ at 1.3~cm wavelength, the effects of scattering substructure have been conclusively detected, apparent as a ${\sim}10$~mJy noise floor on long baselines \citep{Gwinn_2014}. At 3~mm, non-zero closure phases (indicative of structural asymmetry) in \sgra\  are consistent with being introduced by scattering rather than being intrinsic to the source  \citep{Ortiz_2016}. At 1.3~mm, non-zero closure phases in \sgra\ have now been measured by the EHT \citep{Fish_2016}; however, these have persistent sign over four years and so they must predominantly reflect intrinsic source structure. Even so, the epoch-to-epoch variations of these closure phases may be dominated by scattering substructure \citep{Broderick_2016}.  For {\it RadioAstron}, scattering can explain apparent brightness temperatures in excess of $10^{14}~{\rm K}$ in 3C\,273 inferred at 18~cm wavelength \citep{Johnson_2016}, and can dominate long-baseline detections between 6 and 92~cm for sources with high brightness temperatures \citep{Johnson_Gwinn_2015}.  For all these reasons, it is essential to have a straightforward and comprehensive framework to understand the effects of scattering on VLBI observations.

Currently, analytic expressions for scattering effects are limited to quantities such as the root-mean-square (RMS) fluctuations in flux density and complex visibility. Consequently, previous efforts have relied on scattering simulations to estimate effects on realistic VLBI observables such as closure phase and closure amplitude. However, these simulations are computationally expensive and make it difficult to analyze scattering effects over a wide range of source and scattering models. Simulations also provide little insight into how to develop scattering mitigation strategies. 

Here, we derive analytical tools to estimate scattering effects by extending a simplified scattering formalism that was developed by \citet{Blandford_Narayan_1985} to understand the scattering of pulsars. This formalism separates large scale (``refractive'') effects of scattering from small scale (``diffractive'') effects, greatly simplifying computations. It enables a wide range of calculations and can even be used to estimate the covariance between different scattering effects \citep[see also][]{Romani_1986}. We extend the formalism in two directions: to interferometry, and to sources with arbitrary intrinsic structure.  We show how to obtain exact estimates of refractive scattering effects on VLBI observables using a Monte Carlo method and we also derive an approximate form for the closure phase fluctuations when refractive fluctuations are small. 

\clearpage

\section{Theoretical Background}

We first summarize the basic prescription for interstellar scattering of radio waves. For a more detailed discussion and review, see \citet{Rickett_1990} or \citet{Narayan_1992}. When possible, our treatment and notation mirror those of \citet{Blandford_Narayan_1985}.

\subsection{Interstellar Scattering}
\label{sec::ISM}

The local index of refraction $n$ of the ionized interstellar medium (ISM) is determined by its plasma frequency, $\nu_{\rm p} = \sqrt{\frac{n_{\rm e} e^2}{\pi m_{\rm e}}} \approx 9.0 \times \sqrt{\frac{n_{\rm e}}{1~{\rm cm}^{-3}}}\,{\rm kHz}$, where $n_{\rm e}$, $e$, and $m_{\rm e}$ are the local electron number density, electron charge, and electron mass \citep{Jackson_1999}. At frequencies $\nu$ significantly higher than $\nu_{\rm p}$, $n \approx 1 - \frac{1}{2} \left( \frac{\nu_{\rm p}}{\nu} \right)^2$. A density fluctuation $\delta n_{\rm e}$ along a path length $dz$ then introduces a corresponding phase change $\delta \phi = -r_{\rm e} \lambda \times dz \times \delta n_{\rm e}$, where $r_{\rm e} = e^2/(m_{\rm e} c^2) \approx 2.8 \times 10^{-13}~{\rm cm}$ is the classical electron radius and $\lambda$ is the wavelength. Consequently, inhomogeneities in the density of the ionized ISM scatter radio waves. Note that the inhomogeneities have the opposite action of conventional lenses because the phase velocity in a plasma increases with density (and is superluminal).

In many cases, the density fluctuations are well-described as a turbulent cascade, injected at scales ${\gsim}100~{\rm AU}$ and dissipated at scales ${\lsim}1000~{\rm km}$ \citep{Armstrong_1995}.  
Moreover, the scattering can often be well-described as being confined to a single thin screen, located a distance $D$ from the observer and $R$ from the source, which only affects the phase of the incident radiation via an additive contribution $\phi(\mathbf{r})$, where $\mathbf{r}$ is a transverse two-dimensional vector on the screen. This screen is typically assumed to be ``frozen,'' so that the scattering evolution is deterministic, depending only on the relative transverse motions of the observer, screen, and source with a characteristic velocity $\mathbf{V}_{\perp}$ \citep{Taylor_1938}.

Geometrical effects of the scattering are then described via the Fresnel scale, $r_{\rm F} \equiv \sqrt{ \frac{ D R }{D + R} \lambdabar }$ (where $\lambdabar \equiv \frac{\lambda}{2\pi}$), while the screen phase statistics are described via the phase decorrelation length, $r_0$, which is the transverse scale on the scattering screen over which the RMS phase difference is 1 radian. Another relevant quantity is the magnification parameter $M=D/R$ of the scattering screen. In the strong-scattering regime, defined by $r_0 \ll r_{\rm F}$, the refractive scale $r_{\rm R} \equiv r_{\rm F}^2/r_0$ is also important, roughly defining the extent of the scattered image of a point source. The present paper is primarily focused on the strong-scattering regime, although our results are also applicable in the weak-scattering regime if the angular size of the unscattered source is larger than $r_{\rm F}/D$ (see \S\ref{sec::RefractiveSteering}).  

In the strong scattering limit, scattering effects are dominated by phase fluctuations on two widely separated scales. ``Diffractive'' scattering arises from small-scale fluctuations, dominated on scales of ${\sim}r_0$. Diffractive effects decorrelate over a small fractional bandwidth, $(r_0/r_{\rm F})^2$, and have a coherence timescale of only ${\sim}r_0/V_{\perp}$ (typically seconds to minutes). Diffractive effects are quenched for a source that has an angular size larger than $r_0/D$, so they are typically only seen in pulsars. In contrast, ``refractive'' scattering arises from large-scale fluctuations, dominated on scales of ${\sim}r_{\rm R}$. Refractive effects have a decorrelation bandwidth of order unity and a coherence timescale of ${\sim}r_{\rm R}/V_{\perp}$ (typically days to months). Refractive effects are only quenched for a source that has an angular size larger than $r_{\rm R}/D$, so they are present in compact AGNs \citep{RCB_1984}. 

\citet{NarayanGoodman89} and \citet{GoodmanNarayan89} showed that there are three distinct averaging regimes for images in the strong-scattering limit: a ``snapshot'' image, an ``average'' image, and an ``ensemble-average'' image. The snapshot image occurs for averaging timescales less than $r_0/V_{\perp}$ and exhibits both diffractive and refractive scintillation. The average image occurs for longer averaging timescales that are still shorter than $r_{\rm R}/V_{\perp}$, and this regime only exhibits refractive scintillation. The ensemble-average image reflects a complete average over a scattering ensemble (or, equivalently, an infinite average over time), and the scattering effects are a deterministic image blurring via convolution of the unscattered image with a scattering kernel. Because AGNs almost always quench diffractive scintillation, only the average and ensemble-average regimes are of significance in most cases. Throughout this paper, we will use subscripts ``ss'', ``a'', and ``ea'' to denote quantities in the snapshot, average, and ensemble-average regimes, respectively.

\subsection{Statistics of the Screen Phase}
\label{sec::ScreenPhaseStatistics}

We now introduce some properties, notation, and results related to the screen phase. We assume that the function $\phi(\mathbf{r})$ defines a Gaussian random field that is statistically homogeneous but not necessarily isotropic. This field is most intuitively characterized by the phase structure function:
\begin{align}
D_{\phi}(\mathbf{r}) \equiv \left \langle \left[ \phi(\mathbf{r}' + \mathbf{r}) - \phi(\mathbf{r}')  \right]^2 \right \rangle.
\end{align}
Here and throughout the paper, $\langle \dots \rangle$ denotes an ensemble average over realizations of the screen phase. In practice, the ensemble-average can be approximated by averaging over time. The phase varies smoothly up to some ``inner scale,'' which is physically associated with the scale on which turbulence is dissipated. Consequently, $D_{\phi}(\mathbf{r}) \propto |\mathbf{r}|^2$ up to this scale, though it need not be isotropic \citep{Tatarskii_1971}. The phase structure function is then described by a power law in an inertial range until saturating at some ``outer scale,'' which is physically associated with the scale on which the turbulence is injected. 

Another important representation is the two-point correlation function of phase:
\begin{align}
C(\mathbf{r}) \equiv \left \langle \phi(\mathbf{r}' + \mathbf{r}) \phi(\mathbf{r}') \right \rangle.
\end{align}
We can also define the power spectrum of the phase fluctuations:
\begin{align}
\label{eq::Q_eq}
Q(\mathbf{q}) &= \frac{1}{\lambdabar^2} \int d^2\mathbf{r}\, C(\mathbf{r}) e^{-i \mathbf{q} \cdot \mathbf{r}}.
\end{align}
Note that $Q(\mathbf{q})$ is dimensionless and is independent of wavelength, and $\lambdabar^2 |\mathbf{q}|^2 Q(\mathbf{q})$ gives the mean-squared phase fluctuations on a spatial scale $2\pi/|\mathbf{q}|$. Because $D_{\phi}({\bf r}) = 2\left[ C({\bf 0}) - C({\bf r}) \right]$, we can also express Eq.~\ref{eq::Q_eq} as $Q( \mathbf{q} ) \equiv -\frac{1}{2\lambdabar^2}\tilde{D}_{\phi}(\mathbf{q})$. In this expression and throughout this paper, we use a tilde to denote a Fourier transform and adopt the convention that
\begin{align}
\tilde{f}(\mathbf{q},\lambdabar) &= \int d^2\mathbf{r}\, f(\mathbf{r},\lambdabar) e^{-i \mathbf{q} \cdot \mathbf{r}}.
\end{align}
For an isotropic power-law structure function, $D_{\phi}(\mathbf{r}) = \left| \mathbf{r}/r_0 \right|^\alpha$, we obtain\footnote{For 3D Kolmogorov turbulence, $\alpha=5/3$. The present paper focuses on ``shallow'' spectra: $0 < \alpha < 2$.}  
\begin{align}
Q(\mathbf{q}) &= 2^{\alpha} \pi \alpha \frac{\Gamma\left(1+\alpha/2\right)}{\Gamma\left(1-\alpha/2\right)} \lambdabar^{-2} r_0^{-\alpha} \left| \mathbf{q} \right|^{-(2+\alpha)}.
\end{align}
These expressions can easily be generalized to include inner and outer scales or anisotropic scattering. One convenient form is, 
\begin{align}
Q(\mathbf{q}) &= 2^{\alpha} \pi \alpha \frac{\Gamma\left(1+\alpha/2\right)}{\Gamma\left(1-\alpha/2\right)} \lambdabar^{-2} \left( r_{0,x} r_{0,y} \right)^{-\alpha/2} \left[ \left(\frac{r_{0,x}}{r_{0,y}}\right) q_{x}^2 + \left(\frac{r_{0,y}}{r_{0,x}}\right) q_{y}^2 + q_{\rm min}^2 \right]^{-(1 + \alpha/2)} e^{-(q/q_{\rm max})^2},
\end{align}
where the phase decoherence scales can now differ in two orthogonal directions, $r_{0,x}$ and $r_{0,y}$, so that the scattering has an axial ratio of $r_{0,x}/r_{0,y}$ (and the smaller of the two determines the major axis of the scattering disk). 

The above results allow us to derive an important computational tool that we will employ throughout the remainder of this paper. Namely, given two complex functions $f_i(\mathbf{r}_i,\lambda_i)$, 
\begin{align}
\label{eq::BN_Reduction}
\nonumber \left \langle \int d^2\mathbf{r}_{1} d^2\mathbf{r}_{2}\, \phi(\mathbf{r}_1,\lambda_1) \phi(\mathbf{r}_2,\lambda_2) f_1(\mathbf{r}_1,\lambda_1) f_2(\mathbf{r}_2,\lambda_2) \right \rangle 
&= \int d^2\mathbf{r}_{1} d^2\mathbf{r}_{2}\, \frac{\lambda_2}{\lambda_1}  \left \langle \phi(\mathbf{r}_1,\lambda_1) \phi(\mathbf{r}_2,\lambda_1) \right \rangle f_1(\mathbf{r}_1,\lambda_1) f_2(\mathbf{r}_2,\lambda_2)\\
\nonumber &= \frac{\lambda_2}{\lambda_1} \int d^2\mathbf{r}_{2} \frac{1}{(2\pi)^2} \left[ \int d^2\mathbf{q}\, \lambdabar_1^2 Q(\mathbf{q}) e^{-i \mathbf{q} \cdot \mathbf{r}_2} \tilde{f}_1(-\mathbf{q},\lambda_1) \right] f_2(\mathbf{r}_2,\lambda_2)\\
&= \frac{\lambda_1 \lambda_2}{(2\pi)^4} \int d^2 \mathbf{q}\, Q(\mathbf{q}) \tilde{f}_1(-\mathbf{q},\lambda_1) \tilde{f}_2(\mathbf{q},\lambda_2).
\end{align}
The second equality follows from the general identify $\int d^2\mathbf{r}\, g_1(\mathbf{r}) g_2(\mathbf{r}) = \frac{1}{(2\pi)^2} \int d^2\mathbf{q}\, \tilde{g}_1(\mathbf{q}) \tilde{g}_2(-\mathbf{q})$. 
Eq.~\ref{eq::BN_Reduction} was given by \citet{Blandford_Narayan_1985} (their Eq.~4.4) for the special case that the functions $f_i$ are real. Eq.~\ref{eq::BN_Reduction} can also easily be extended to the useful cases where one or both of the functions $f_i$ are conjugated:
\begin{align}
\label{eq::BN_Reduction_Conj}
\left \langle \int d^2\mathbf{r}_{1} d^2\mathbf{r}_{2}\, \phi(\mathbf{r}_1,\lambda_1) \phi(\mathbf{r}_2,\lambda_2) f_1(\mathbf{r}_1,\lambda_1) f_2^\ast(\mathbf{r}_2,\lambda_2) \right \rangle &= \frac{\lambda_1 \lambda_2}{(2\pi)^4} \int d^2\mathbf{q}\, Q(\mathbf{q}) \tilde{f}_1(\mathbf{q},\lambda_1) \tilde{f}_2^\ast(\mathbf{q},\lambda_2),\\
\nonumber \left \langle \int d^2\mathbf{r}_{1} d^2\mathbf{r}_{2}\, \phi(\mathbf{r}_1,\lambda_1) \phi(\mathbf{r}_2,\lambda_2) f_1^\ast(\mathbf{r}_1,\lambda_1) f_2^\ast(\mathbf{r}_2,\lambda_2) \right \rangle &= \frac{\lambda_1 \lambda_2}{(2\pi)^4} \int d^2\mathbf{q}\, Q(\mathbf{q}) \tilde{f}_1^\ast(\mathbf{q},\lambda_1) \tilde{f}_2^\ast(-\mathbf{q},\lambda_2).
\end{align}

\subsection{Separation of Diffractive and Refractive Scattering}
\label{sec::RefractiveSteering}

The framework of \citet{Blandford_Narayan_1985} greatly simplifies the scattering by separating diffractive and refractive effects. We simply sketch the key ideas here but provide formal justification for this separation of scales in Appendix~\ref{sec::ScaleSeparation}. 

Diffractive effects reflect stochastic fluctuations on small scales, so their influence on an image can be approximated through an ensemble average. The ensemble average effect is a convolution of the intensity distribution of the unscattered image, $I_{\rm src}(\mathbf{r})$, with a deterministic kernel, resulting in the ensemble-average image $I_{\rm ea}(\mathbf{r})$, which is a ``blurred'' version of the unscattered image \citep[for details and discussion, see][]{Fish_2014}. We again emphasize that $\mathbf{r}$ is a transverse coordinate at the distance $D$ of the scattering screen, so the corresponding angular coordinates for the images are $\boldsymbol{\theta} = \mathbf{r}/D$.

Refractive effects arise from fluctuations on much larger scales, and their stochastic effects persist in the average-image regime. Gradients in the refractive component of the phase screen, $\phi_{\rm r}(\mathbf{r})$, steer and distort brightness elements of the unscattered image while preserving the surface brightness \citep{Born_Wolf}. The Fresnel scale, $r_{\rm F}$, determines how these phase gradients affect the average image, such that the observed brightness received at some location $\mathbf{x}=0$ in the observing plane can be expressed as
\begin{align}
\label{eq::RefractiveSteering_Prelim}
I_{\rm a}(\mathbf{r}) &\approx I_{\rm ea}(\mathbf{r} + r_{\rm F}^2 \nabla \phi_{\rm r}(\mathbf{r})).
\end{align}
Here, $\nabla$ denotes a transverse (two-dimensional) spatial gradient operator. On diffractive scales, the phase gradient is $\left| \nabla \phi(\mathbf{r})) \right| \sim 1/r_0$, whereas the gradient from phase fluctuations on a larger, refractive scale $L$ is $\left| \nabla \phi(\mathbf{r})) \right| \sim \sqrt{D_{\phi}(L)}/L \sim L^{\alpha/2-1}/r_0^{\alpha/2}$. Consequently, refractive steering angles are smaller than diffractive steering angles by a factor of ${\sim}(r_0/r_{\rm R})^{1-\alpha/2} = (r_0/r_{\rm F})^{2-\alpha}$, and this factor roughly determines the relative strength of refractive and diffractive effects for metrics such as flux modulation \citep{Narayan_1992}. Moreover, because the refractive steering is smaller than the diffractive blurring, we can approximate the average image by expanding to leading order:
\begin{align}
\label{eq::RefractiveSteering}
I_{\rm a}(\mathbf{r}) &\approx I_{\rm ea}(\mathbf{r}) + r_{\rm F}^2 \left[\nabla \phi_{\rm r}(\mathbf{r})\right] \cdot \left[ \nabla I_{\rm ea}(\mathbf{r}) \right].
\end{align}
Note that this approximation is valid even in the weak-scattering regime ($r_0 > r_{\rm F}$) if the intrinsic source does not have significant spatial variations in its structure on scales finer than the typical scattering angle (i.e., if the interferometric visibility of the unscattered source is negligible on baselines that resolve the typical scattering angle). For this reason, there is a unified description of refractive scattering effects in the strong- and weak-scattering regimes for an extended source \citep[see, e.g.,][]{Coles_1987}.

We are interested in describing VLBI observations in the average-image regime, for which diffractive scintillation is replaced by its ensemble-average effects. Moreover, because we are interested in VLBI observations that resolve intrinsic source structure, we will assume that the source is sufficiently extended to quench diffractive scintillation, so no additional averaging in time is required (in \S\ref{sec::Interferometry}, we will show how an extended source acts to filter contributions from wavenumbers $\mathbf{q}$ on diffractive scales). For these reasons, we can substitute $\phi(\mathbf{r})$ for $\phi_{\rm r}(\mathbf{r})$ in Eq.~\ref{eq::RefractiveSteering}, and we will use this replacement throughout the remainder of the paper for simplicity.

\subsection{Interferometry of Scattered Sources}
\label{sec::Interferometry}

Using the approximate form for the scattered image given in Eq.~\ref{eq::RefractiveSteering} and the statistical properties of the screen phase $\phi$ derived in \S\ref{sec::ScreenPhaseStatistics}, it is straightforward to estimate how refractive scattering affects interferometric visibilities. Recall that the average image in Eq.~\ref{eq::RefractiveSteering} is defined at the location of the scattering screen, located a distance $D$ from the observer. The van~Cittert-Zernike Theorem\footnote{The van~Cittert-Zernike Theorem only strictly applies when a source is spatially incoherent. Scattering introduces spatial coherence, but because the correlation length is $r_0$, the source appears spatially incoherent to interferometric baselines $|\mathbf{b}| \ll \lambda (r_0/D)^{-1} \sim r_{\rm R}$, as holds for all cases of immediate interest (see Appendix~\ref{sec::ScaleSeparation} for details).} then gives the interferometric visibility on a baseline $\mathbf{b}$ \citep{TMS}:
\begin{align}
\label{eq::Vavg}
\nonumber V_{\rm a}(\mathbf{b}) &= \int d^2\mathbf{r}\, I_{\rm a}(\mathbf{r}) e^{-i \mathbf{r} \cdot \mathbf{b}/(D \lambdabar)}\\
\nonumber &\approx V_{\rm ea}(\mathbf{b}) + r_{\rm F}^2 \int d^2\mathbf{r}\,  e^{-i \mathbf{r} \cdot \mathbf{b}/(D \lambdabar)}\nabla \phi(\mathbf{r}) \cdot \nabla I_{\rm ea}(\mathbf{r})\\
\nonumber &= V_{\rm ea}(\mathbf{b}) + r_{\rm F}^2 \int d^2\mathbf{r}\,  e^{-i \mathbf{r} \cdot \mathbf{b}/(D \lambdabar)} \left[ \frac{i}{D\lambdabar} \mathbf{b} \cdot \nabla I_{\rm ea}(\mathbf{r}) - \nabla^2 I_{\rm ea}(\mathbf{r})  \right] \phi(\mathbf{r})\\
&\equiv V_{\rm ea}(\mathbf{b}) + \int d^2\mathbf{r}\,  f_{\rm V}(\mathbf{r};\mathbf{b},\lambda) \phi(\mathbf{r}).
\end{align}
The third line was obtained by integrating by parts (we assume that $I_{\rm ea}(\mathbf{r})$ is restricted to a finite domain, so the boundary term vanishes), and we have defined
\begin{align}
\label{eq::fV_def}
f_{\rm V}(\mathbf{r};\mathbf{b},\lambda) &\equiv r_{\rm F}^2 e^{-i \mathbf{r} \cdot \mathbf{b}/(D \lambdabar)} \left[ \frac{i}{D\lambdabar} \mathbf{b} \cdot \nabla I_{\rm ea}(\mathbf{r}) - \nabla^2 I_{\rm ea}(\mathbf{r}) \right].
\end{align}
Note that $f_{\rm V}$ depends on the ensemble-average image. This dependence is important when comparing the refractive noise at different frequencies (with different intrinsic structure), when studying the refractive noise for an image that varies more rapidly than the scattering (as may be the case for \sgra), or when comparing the refractive noise for images corresponding to different Stokes parameters. When convenient, we will write quantities in terms of the ensemble-average visibility, $V_{\rm ea}(\mathbf{b})$, which is easily related to the interferometric visibility of the unscattered source, $V_{\rm src}(\mathbf{b})$, through the convolution action of the scattering in this regime \citep[e.g.,][]{Coles_1987}:
\begin{align}
\label{eq::Vea}
V_{\rm ea}(\mathbf{b}) = V_{\rm src}(\mathbf{b}) e^{-\frac{1}{2} D_{\phi}\left( \frac{\mathbf{b}}{1+M} \right)}. 
\end{align}

From Eq.~\ref{eq::Vavg}, it is apparent that refractive noise on interferometric visibilities is Gaussian because it is simply a weighted sum of (correlated) Gaussian random variables, $\phi(\mathbf{r})$. Because the refractive noise is Gaussian, it is completely characterized by its spatial covariance.  However, the real and imaginary parts of the Gaussian noise may have different standard deviations and different spatial covariance structure; for example, on the baseline $\mathbf{b}=\mathbf{0}$, the refractive noise is purely real since the zero-baseline visibility is real (and equal to the total flux density of the source). On long baselines (i.e., those baselines that completely resolve the ensemble-average image) the refractive noise is drawn from a circular complex Gaussian distribution. 

To calculate the full spatial covariance of refractive noise, we must therefore derive expressions for the real and imaginary parts of the noise separately. These are simplified by the observation that $f_{\rm V}$ obeys the necessary conjugation symmetry for interferometric visibilities: $f_{\rm V}^\ast(\mathbf{r};\mathbf{b},\lambda) = f_{\rm V}(\mathbf{r};-\mathbf{b},\lambda)$ (see Eq.~\ref{eq::Vavg}). Consequently,
\begin{align}
\label{eq::fVreim}
f_{\rm V,re}(\mathbf{r};\mathbf{b},\lambda) &\equiv \mathrm{Re}\left[ f_{\rm V}(\mathbf{r};\mathbf{b},\lambda) \right]\\
\nonumber &= \frac{1}{2} \left[ f_{\rm V}(\mathbf{r};\mathbf{b},\lambda) + f_{\rm V}(\mathbf{r};-\mathbf{b},\lambda) \right] \\
\nonumber &= r_{\rm F}^2 \left\{ -\cos\left[ \mathbf{r} \cdot \mathbf{b}/(D \lambdabar) \right] \nabla^2 I_{\rm ea}(\mathbf{r}) + \sin\left[ \mathbf{r} \cdot \mathbf{b}/(D \lambdabar) \right] \frac{1}{D\lambdabar} \mathbf{b} \cdot \nabla I_{\rm ea}(\mathbf{r}) \right\}, \\ 
\nonumber f_{\rm V,im}(\mathbf{r};\mathbf{b},\lambda) &\equiv \mathrm{Im}\left[ f_{\rm V}(\mathbf{r};\mathbf{b},\lambda) \right]\\
\nonumber &= \frac{1}{2i} \left[ f_{\rm V}(\mathbf{r};\mathbf{b},\lambda) - f_{\rm V}(\mathbf{r};-\mathbf{b},\lambda) \right] \\
\nonumber &= r_{\rm F}^2 \left\{+\sin\left[ \mathbf{r} \cdot \mathbf{b}/(D \lambdabar) \right] \nabla^2 I_{\rm ea}(\mathbf{r}) + \cos\left[ \mathbf{r} \cdot \mathbf{b}/(D \lambdabar) \right] \frac{1}{D\lambdabar} \mathbf{b} \cdot \nabla I_{\rm ea}(\mathbf{r}) \right\}.
\end{align}

The corresponding Fourier conjugate quantities are then
\begin{align}
\label{eq::fVtilde}
\tilde{f}_{\rm V}(\mathbf{q};\mathbf{b},\lambda) &= \int d^2\mathbf{r}\, f_{\rm V}(\mathbf{r};\mathbf{b},\lambda) e^{-i \mathbf{q}\cdot \mathbf{r}}\\
\nonumber &= r_{\rm F}^2 \int d^2\mathbf{r}\, e^{-i \mathbf{r} \cdot \left[\mathbf{q} + \mathbf{\mathbf{b}}/(D\lambdabar) \right]} \left[ \frac{i}{D\lambdabar} \mathbf{b} \cdot \nabla I_{\rm ea}(\mathbf{r}) - \nabla^2 I_{\rm ea}(\mathbf{r})  \right]\\
\nonumber &= r_{\rm F}^2 \left[ \mathbf{q} + \mathbf{b}/(D\lambdabar) \right] \cdot \left\{ \left[ \mathbf{q} + \mathbf{b}/(D\lambdabar) \right] - \frac{\mathbf{b}}{D \lambdabar} \right\} V_{\rm ea}\left( D\lambdabar\mathbf{q}  + \mathbf{b} \right)\\
\nonumber &= r_{\rm F}^2 \mathbf{q} \cdot \left[ \mathbf{q} + (1+M)^{-1} r_{\rm F}^{-2} \mathbf{b} \right] V_{\rm ea}\left( (1+M)r_{\rm F}^2 \mathbf{q}  + \mathbf{b} \right),\\
\nonumber \tilde{f}_{\rm V,re}(\mathbf{q};\mathbf{b},\lambda) &= \frac{1}{2} \left[ \tilde{f}_{\rm V}(\mathbf{q};\mathbf{b},\lambda) + \tilde{f}_{\rm V}(\mathbf{q};-\mathbf{b},\lambda) \right],\\
\nonumber \tilde{f}_{\rm V,im}(\mathbf{q};\mathbf{b},\lambda) &= \frac{1}{2i} \left[ \tilde{f}_{\rm V}(\mathbf{q};\mathbf{b},\lambda) - \tilde{f}_{\rm V}(\mathbf{q};-\mathbf{b},\lambda) \right],
\end{align}
where we have again integrated by parts, dropping the boundary term because we expect a source image with finite domain, and we have made the substitution $D \lambdabar = (1 + M) r_{\rm F}^2$. 

When coupled with Eq.~\ref{eq::BN_Reduction}, Eq.~\ref{eq::fVtilde} shows how the source, scattering, and observing baseline all act to filter different wavenumbers $\mathbf{q}$ of the screen phase. For instance, on long baselines the ensemble-average visibility filters wavenumbers that are significantly different than $-(1+M)^{-1} r_{\rm F}^{-2}\mathbf{b} = -\mathbf{b}/(\lambdabar D)$. Thus, the contributing wavenumbers are those with a corresponding angular scale, $\left(2\pi/\mathbf{q}\right)/D$, that is matched to the vector baseline resolution, $\lambda/\mathbf{b}$. On short baselines, the ensemble-average visibility filters wavenumbers with $|\mathbf{q}| \gsim r_0/r_{\rm F}^2 = r_{\rm R}^{-1}$. In both cases, this term filters the wavenumbers that produce diffractive scintillation because they have $|\mathbf{q}| \sim r_0^{-1} \gg r_{\rm R}^{-1}$, and the ensemble-average visibility falls to zero for baselines $|\mathbf{b}| \gg r_0$ (see Eq.~\ref{eq::Vea}). The term $r_{\rm F}^2 \mathbf{q} \cdot \left[ \mathbf{q} + (1+M)^{-1} r_{\rm F}^{-2} \mathbf{b} \right]$ filters power at low wavenumbers because of wrapping of the phase from geometrical path length. The characteristic cutoff for this term is $|\mathbf{q}| \lsim 1/r_{\rm F}$, so this term is called the Fresnel filter \citep{Cronyn_1972}. The Fresnel filter regulates power on refractive scales, so the strength of refractive scintillation falls as the scales $r_0$, $r_{\rm F}$, and $r_{\rm R}$ become more widely separated.

\section{Signatures of Refractive Noise}

From the results derived in \S\ref{sec::Interferometry}, it is straightforward to compute the effects of refractive scattering on interferometric observables. We begin by calculating the RMS fluctuations of interferometric visibilities, showing that our formalism exactly reproduces the results of \citet{GoodmanNarayan89} and \citet{Johnson_Gwinn_2015} but requires significantly less effort than in their approach. We then show how to calculate arbitrary observables to any desired accuracy using a Monte Carlo framework, and we derive approximations for the refractive fluctuations in closure phase. We conclude this section by discussing the coherence timescale of refractive noise.

\subsection{Refractive Fluctuations of the Complex Interferometric Visibility}
\label{sec::sigma_ref}

To make contact with previous results and illustrate our computational methodology, we will first estimate the RMS fluctuation of the complex visibility introduced by refractive substructure on a fixed baseline $\mathbf{b}$. Combining Eqs.~\ref{eq::BN_Reduction_Conj}, \ref{eq::Vavg}, and \ref{eq::fVtilde}, we obtain
\begin{align}
\label{eq::sigmaref_Q}
\sigma_{\rm ref}^2(\mathbf{b}) \equiv \left \langle \left| \Delta V_{\rm a}(\mathbf{b}) \right|^2 \right \rangle &= \frac{\lambda^2}{(2\pi)^4} \int d^2\mathbf{q}\, \left| \tilde{f}_{\rm V}(\mathbf{q};\mathbf{b},\lambda)  \right|^2 Q(\mathbf{q})\\
\nonumber &= \frac{\lambda^2 r_{\rm F}^4}{(2\pi)^4} \int d^2\mathbf{q}\,  \left( \mathbf{q} \cdot \left[ \mathbf{q} + \mathbf{b}/(D\lambdabar) \right] \right)^2 \left|V_{\rm ea}\left( D\lambdabar\mathbf{q}  + \mathbf{b} \right) \right|^2 Q(\mathbf{q}), 
\end{align}
where $\Delta V_{\rm a}(\mathbf{b}) \equiv V_{\rm a}(\mathbf{b}) - \langle V_{\rm a}(\mathbf{b}) \rangle = V_{\rm a}(\mathbf{b}) - V_{\rm ea}(\mathbf{b})$. 
Performing the substitutions $Q(\mathbf{q}) = -\frac{1}{2\lambdabar^2} \tilde{D}_\phi(\mathbf{q})$, $\mathbf{y} = \left( \mathbf{q} + \mathbf{b}/(D\lambdabar) \right) r_{\rm F}^2$, and $\lambdabar D = (1 + M) r_{\rm F}^2$ then gives
\begin{align}
\label{eq::sigmaref}
\sigma_{\rm ref}^2(\mathbf{b}) = -\frac{1}{8 \pi^2 r_{\rm F}^8} \int d^2\mathbf{y}\, \left( \mathbf{y} \cdot \left[ \mathbf{y} - (1+M)^{-1} \mathbf{b}  \right] \right)^2 \left|V_{\rm ea}\left( (1+M)\mathbf{y} \right) \right|^2 \tilde{D}_{\phi}\left(r_{\rm F}^{-2} \left[\mathbf{y} - (1+M)^{-1}\mathbf{b}\right] \right).
\end{align}
Apart from an opposite convention for the sign of $\mathbf{b}$ and choice of units for the argument of $\tilde{D}_{\phi}$, Eq.~\ref{eq::sigmaref} exactly matches Eq.~32 of \citet{Johnson_Gwinn_2015}, which was derived using a more general formulation \citep[based on][]{GoodmanNarayan89}. This agreement strongly reinforces the utility of the present approach and demonstrates its capability for quickly estimating results that formerly required arduous derivations. \citet{Cronyn_1972} considered visibility scintillation in the weak-scattering regime and derived a result that is similar to Eq.~\ref{eq::sigmaref_Q} \citep[see also][]{Vedantham_2015}.

Although Eq.~\ref{eq::sigmaref} requires numerical integration in the most general case, it can be evaluated in closed form for a Gaussian source in the limit of long baseline (specifically, a baseline that resolves the ensemble-average image). For isotropic scattering and a circular Gaussian intrinsic source with FWHM $\theta_{\rm src}$, we obtain
\begin{align}
\label{eq::sigmaref_approx}
\sigma_{\rm ref}(\mathbf{b}) &\approx \sqrt{\frac{\Gamma(4/\alpha)}{2^{2-\alpha}}\frac{\Gamma\left(1+\frac{\alpha}{2}\right)}{\Gamma\left(1-\frac{\alpha}{2}\right)}} \left( \frac{r_0}{r_{\rm F}} \right)^{2-\alpha} \left(\frac{\left| \mathbf{b} \right|}{\left(1+M\right)r_0}\right)^{-\frac{\alpha}{2}} \left( \frac{\theta_{\rm scatt}}{\theta_{\rm img}}\right)^2,
\end{align}
where $\theta_{\rm scatt} \approx \frac{\sqrt{2\ln{2}}}{\pi}\frac{\lambda}{(1+D/R)r_0}$ is the scattered angular size of a point source and $\theta_{\rm img} \approx \sqrt{\theta_{\rm src}^2 + \theta_{\rm scatt}^2}$ is the ensemble-average angular size. In the limit of a small intrinsic source, $\theta_{\rm img} \rightarrow \theta_{\rm scatt}$, Eq.~\ref{eq::sigmaref_approx} reproduces the result of \citet{GoodmanNarayan89} (their Eq.~5.1.2).

\begin{figure}[t]
\centering
\includegraphics[width=\textwidth]{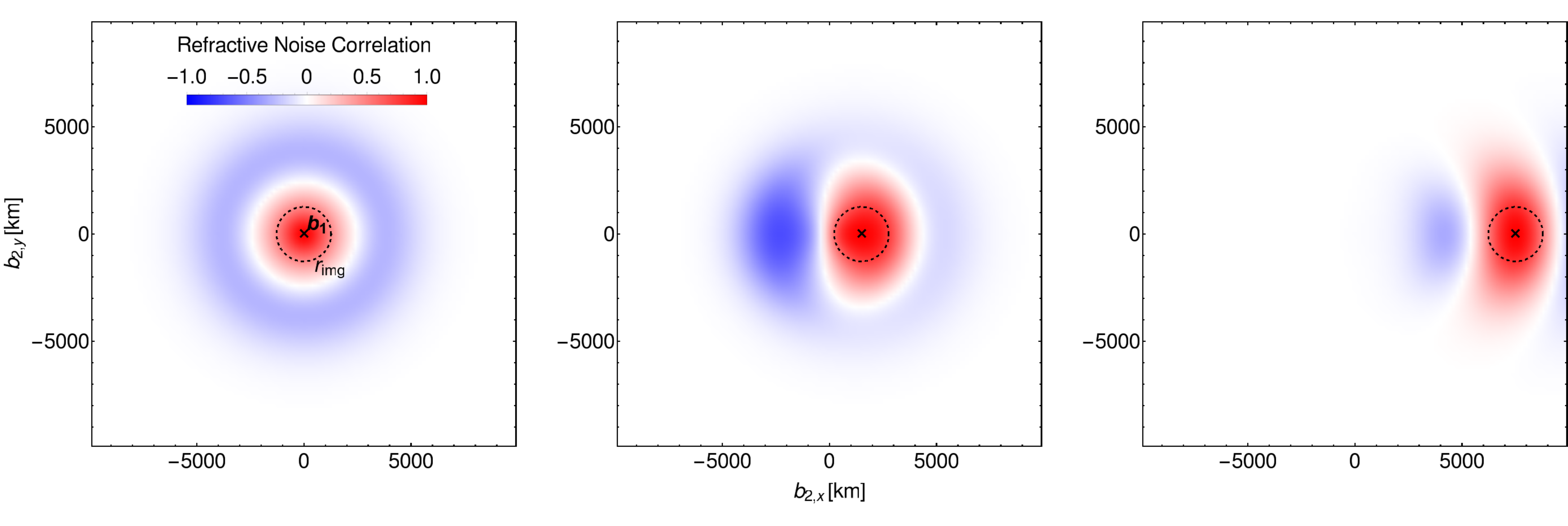}
\caption
{ 
Examples showing the correlation $\left \langle \Delta V_{\rm a}\left(\mathbf{b}_1\right) \Delta V_{\rm a}^\ast\left(\mathbf{b}_2\right) \right \rangle/\left[\sigma_{\rm ref}(\mathbf{b}_1) \sigma_{\rm ref}(\mathbf{b}_2)\right]$ between refractive noise on a fixed interferometric baseline $\mathbf{b}_1$ (indicated by a cross) and refractive noise on another arbitrary baseline $\mathbf{b}_2$. The source (FWHM: $130~\mu{\rm as}$) and scattering (FWHM: $160~\mu{\rm as}$) are both taken to be isotropic and correspond roughly to the major axis of \sgra\ at 3.5~mm wavelength \citep[see][]{Ortiz_2016}. Three choices of $\mathbf{b}_1$ are shown: a zero-baseline (measuring the total flux density), a short baseline (which partially resolves the source), and a long baseline (which entirely resolves the ensemble-average image). As expected, the refractive noise is correlated with a scale of approximately $r_{\rm img}$, which is the baseline length at which the ensemble-average visibility amplitude falls to $1/\sqrt{e}$ of its zero-baseline value. Note that there is also a region of anti-correlation before the noise becomes uncorrelated on baselines that are significantly distant from $\mathbf{b}_1$. Also, note that the correlation in this case is real because the unscattered image is point symmetric.
}
\label{fig::RefractiveNoise_Correlation}
\end{figure}

\begin{figure}[t]
\centering
\includegraphics[width=\textwidth]{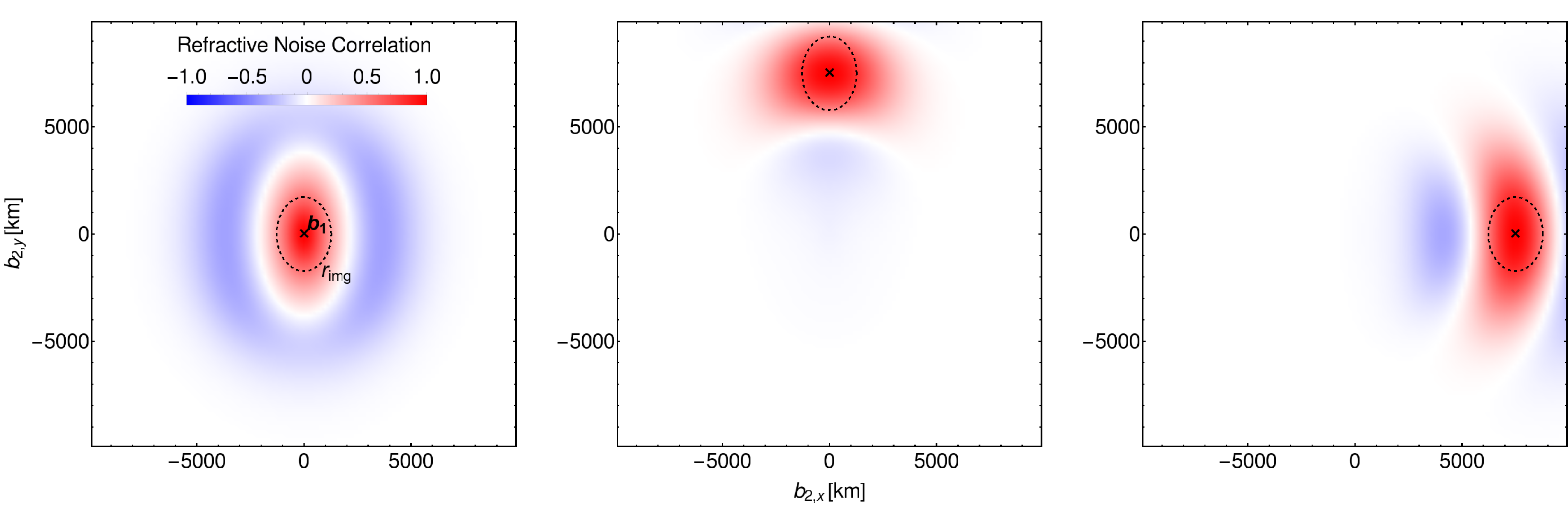}\vspace{-0.3cm}
\caption
{ 
Examples showing the correlation between refractive noise on a fixed interferometric baseline $\mathbf{b}_1$ (indicated by a cross) and another arbitrary baseline $\mathbf{b}_2$, as in Figure~\ref{fig::RefractiveNoise_Correlation} but here with anisotropic scattering. The source (FWHM: $130~\mu{\rm as}$) and scattering (FWHM: $160~\mu{\rm as} \times 78~\mu{\rm as}$) again correspond roughly to \sgra\ at 3.5~mm wavelength, but we have set the major axis of the scattering to be East-West for clarity. Three choices of $\mathbf{b}_1$ are shown: a zero-baseline (left) measuring the total flux density, and two long baselines located along the minor (center) and major (right) axes of the scattering disk. As in Figure~\ref{fig::RefractiveNoise_Correlation}, the refractive noise is correlated with a scale of approximately $r_{\rm img}$, which is now anisotropic.
}
\label{fig::RefractiveNoise_Correlation_Anisotropic}
\end{figure}

We can similarly calculate the covariance of the visibility noise among different baselines, the covariance of the real or imaginary parts of visibility, or the covariance for multiple wavelengths or different underlying images (e.g., among different polarizations). Importantly, none of these was straightforward to estimate in the formulation of \citet{GoodmanNarayan89} or \citet{Johnson_Gwinn_2015}. For example,
\begin{align}
\label{eq::ComplexCovariance}
\left \langle \Delta V_{\rm a}\left(\mathbf{b}_1\right) \Delta V_{\rm a}^\ast\left(\mathbf{b}_2\right) \right \rangle &= \frac{\lambda^2}{(2\pi)^4} \int d^2\mathbf{q}\, \tilde{f}_{\rm V}(\mathbf{q};\mathbf{b}_1,\lambda) \tilde{f}_{\rm V}^\ast(\mathbf{q};\mathbf{b}_2,\lambda) Q(\mathbf{q})\\
\nonumber &= \frac{\lambda^2 r_{\rm F}^4}{(2\pi)^4} \int d^2\mathbf{q}\, \left(\mathbf{q} \cdot \left[ \mathbf{q} + (1+M)^{-1} r_{\rm F}^{-2} \mathbf{b}_1 \right]\right) \left(\mathbf{q} \cdot \left[ \mathbf{q} + (1+M)^{-1} r_{\rm F}^{-2} \mathbf{b}_2 \right]\right)\\
\nonumber &\qquad \qquad \qquad \times V_{\rm ea}\left( (1+M)r_{\rm F}^2 \mathbf{q}  + \mathbf{b}_1 \right) V_{\rm ea}^\ast\left( (1+M)r_{\rm F}^2 \mathbf{q}  + \mathbf{b}_2 \right) Q(\mathbf{q}).
\end{align}
Although this covariance is complex, it is real when $V_{\rm ea}(\mathbf{b}) \in \mathbb{R}$ for all baselines, a condition that is met in the special case of a point-symmetric source. Baselines that are separated by a distance less than ${\sim}r_{\rm img}$ will have correlated refractive fluctuations, where $r_{\rm img}$ is the baseline length needed to resolve the ensemble-average image, as can be derived using more general arguments \citep[see \S3.2 of][]{Johnson_Gwinn_2015}. Note that $r_{\rm img}$ depends on the baseline orientation if the image is anisotropic. Figure~\ref{fig::RefractiveNoise_Correlation} shows the refractive noise correlation structure for isotropic scattering, and Figure~\ref{fig::RefractiveNoise_Correlation_Anisotropic} shows how anisotropic scattering affects the correlation structure.

\subsection{A General Monte Carlo Procedure to Estimate Refractive Effects on Interferometric Observables}
\label{sec::MC}

We now describe a general approach to estimate refractive effects on arbitrary interferometric observables. Although this approach does not produce closed-form expressions, it can easily be computed to any desired accuracy and is not limited to cases in which the refractive noise on a given measurable is small. 

To begin, recall that the refractive fluctuations of a complex visibility about its ensemble-average value correspond to a zero-mean Gaussian random process (see \S\ref{sec::Interferometry}). Consequently, the statistical properties of refractive noise on a set of visibilities (possibly with different sampled times and even different underlying images) are entirely described by the corresponding set of ensemble-average visibilities on those baselines $\{ V_{{\rm ea},i} \}$ and the covariance matrix $\boldsymbol{\Sigma}$ for the set of refractive fluctuations $\{ {\rm Re}\left( \Delta V_{{\rm a},i}\right), {\rm Im}\left( \Delta V_{{\rm a},i}\right) \}$. Explicitly, the probability density function (PDF) of $N$ visibilities is
\begin{align}
\label{eq::PDF}
P(\mathbf{V} \equiv \{ {\rm Re}\left(V_1\right), {\rm Im}\left(V_1\right), \ldots, {\rm Re}\left(V_N\right), {\rm Im}\left(V_N\right) \}) &= \frac{1}{(2\pi)^N \sqrt{|\boldsymbol{\Sigma}|}} \exp\left[ -\frac{1}{2} \left( \mathbf{V} - \mathbf{V}_{\rm ea} \right)^\intercal \boldsymbol{\Sigma}^{-1} \left( \mathbf{V} - \mathbf{V}_{\rm ea} \right) \right].
\end{align}
Each element of the covariance matrix $\boldsymbol{\Sigma}$ must be computed using Eqs.~\ref{eq::BN_Reduction} and \ref{eq::fVtilde}. For example,
\begin{align}
\label{eq::ReRe}
\left \langle {\rm Re}\left[\Delta V_{\rm a}\left(\mathbf{b}_1;\lambda_1 \right)\right] {\rm Re}\left[ \Delta V_{\rm a}\left(\mathbf{b}_2; \lambda_2\right) \right] \right \rangle &= \frac{\lambda_1 \lambda_2}{(2\pi)^4} \int d^2\mathbf{q}\, \tilde{f}_{\rm V,re}(-\mathbf{q};\mathbf{b}_1,\lambda_1) \tilde{f}_{\rm V,re}(\mathbf{q};\mathbf{b}_2,\lambda_2) Q(\mathbf{q}).
\end{align}
This equation is similar to Eq.~\ref{eq::ComplexCovariance}, which instead gave the covariance of the complex visibility.

To evaluate the refractive fluctuations of any quantity derived from a discrete set of visibilities, one can then use a Monte Carlo method by generating complex Gaussian random variables with the appropriate means (the ensemble average) and covariance matrix (the refractive noise).\footnote{Many software packages can generate samples from the multivariate normal distribution for a specified mean and covariance matrix \citep[e.g., {\tt random.multivariate\_normal} in {\tt numpy};][]{SciPy}.} For example, to generate an ensemble of closure phase measurements (see \S\ref{sec::Closure_Phase} below) on a particular triangle requires repeated sampling of $N=3$ complex Gaussian random variables after calculating their $6\times 6$ covariance matrix. To construct an ensemble of closure amplitude realizations on a particular quadrangle requires $N=4$ complex Gaussian random variables and an $8\times 8$ covariance matrix. One could also numerically integrate the desired quantity (e.g., $\mathrm{arg}(V_1 V_2 V_3)$ for closure phase) over the PDF specified by Eq.~\ref{eq::PDF} to estimate properties such as the closure phase variance from scattering. We again emphasize that this approach can be used to calculate the covariance among different images, as is needed to generate an ensemble of measurements for different polarizations, or at different observing wavelengths.

\subsection{Refractive Fluctuations of Closure Phase}
\label{sec::Closure_Phase}
A fundamental interferometric observable is closure phase: the phase of a directed product of three complex visibilities for a closed triangle of baselines (i.e., the phase of the ``bispectrum'') \citep{Rogers_1974,TMS}. Because phase information in VLBI is typically only accessible through closure phase, interferometric imaging algorithms often operate directly on closure phases or even on the bispectrum \citep[see, e.g.,][]{Buscher_1994,Baron_2010,Bouman_2015,Chael_2016}. Consequently, understanding how refractive scattering affects closure phase is critical for VLBI imaging efforts. 

To proceed, we will write the bispectrum $\mathcal{B}$ on a fixed baseline triangle $\left\{ \mathbf{b}_1, \mathbf{b}_2, \mathbf{b}_3 \right\}$ in simplified notation:
\begin{align}
\mathcal{B} = \left( V_1 + \Delta V_1 \right) \left( V_2 + \Delta V_2 \right) \left( V_3 + \Delta V_3 \right),
\end{align}
where $V_i$ denotes an ensemble-average visibility and $\Delta V_i$ denotes the refractive noise for a particular realization of the scattering. To estimate the fluctuations of the phase of the bispectrum, we define a normalized bispectrum:
\begin{align}
\hat{\mathcal{B}} \equiv \mathcal{B}/\mathcal{B}_{\rm ea} =  \left( 1 + \frac{\Delta V_1}{V_1} \right) \left( 1 + \frac{\Delta V_2}{V_2} \right) \left( 1 + \frac{\Delta V_3}{V_3} \right),
\end{align}
where $\mathcal{B}_{\rm ea} \equiv V_1 V_2 V_3$ is the bispectrum of the ensemble-average image. 
When $|\Delta V_i/V_i| \ll 1$ for each $i$, the RMS closure phase fluctuation is then $\sigma_{\rm CP} \approx \sqrt{\left \langle {\rm Im}\left( \hat{\mathcal{B}} \right)^2 \right \rangle}~{\rm radians}$. Keeping only terms that are quadratic in $\Delta V_i/V_i$, we obtain:
\begin{align}
{\rm Im}\left( \hat{\mathcal{B}} \right)^2 = -\frac{1}{4}\left[ \hat{\mathcal{B}} - \hat{\mathcal{B}}^\ast \right]^2 \approx -\frac{1}{4} \left[ \sum_{i=1}^3  \left( \frac{\Delta V_i}{V_i} - \frac{\Delta V_i^\ast}{V_i^\ast} \right) \right]^2.  
\end{align}
Taking the ensemble-average of this expression (via Eqs.~\ref{eq::BN_Reduction}, \ref{eq::BN_Reduction_Conj}, and \ref{eq::Vavg}) then yields
\begin{align}
\label{eq::ClosurePhase}
\sigma_{\rm CP}^2(\{ \mathbf{b}_\ell \}) &\approx -\frac{1}{4} \left \langle \left[ \sum_{i=1}^3  \left( \frac{\Delta V_i}{V_i} - \frac{\Delta V_i^\ast}{V_i^\ast} \right) \right]^2 \right \rangle \\
\nonumber &= -\frac{1}{4} \left \langle \sum_{i,j=1}^3  \left( \frac{\Delta V_i}{V_i} \frac{\Delta V_j}{V_j} + \frac{\Delta V_i^\ast}{V_i^\ast} \frac{\Delta V_j^\ast}{V_j^\ast} -  \frac{\Delta V_i^\ast}{V_i^\ast} \frac{\Delta V_j}{V_j} -  \frac{\Delta V_i}{V_i} \frac{\Delta V_j^\ast}{V_j^\ast} \right) \right \rangle \\
\nonumber &= -\frac{1}{4} \frac{\lambda^2}{(2\pi)^4} \int d^2 \mathbf{q}\, Q(\mathbf{q}) \sum_{i,j=1}^3  \left[
\left( \frac{\tilde{f}_{\rm V}(\mathbf{q};\mathbf{b}_i) \tilde{f}_{\rm V}(-\mathbf{q};\mathbf{b}_j)}{V_{\rm ea}(\mathbf{b}_i) V_{\rm ea}(\mathbf{b}_j)} 
+ \frac{\tilde{f}^\ast_{\rm V}(\mathbf{q};\mathbf{b}_i) \tilde{f}^\ast_{\rm V}(-\mathbf{q};\mathbf{b}_j)}{V_{\rm ea}^\ast(\mathbf{b}_i) V_{\rm ea}^\ast(\mathbf{b}_j)}  \right)
\right. \\
\nonumber & \left. \hspace{0.3\linewidth} 
- \left( \frac{\tilde{f}^\ast_{\rm V}(\mathbf{q};\mathbf{b}_i) \tilde{f}_{\rm V}(\mathbf{q};\mathbf{b}_j)}{V_{\rm ea}^\ast(\mathbf{b}_i) V_{\rm ea}(\mathbf{b}_j)} 
+ \frac{\tilde{f}_{\rm V}(\mathbf{q};\mathbf{b}_i) \tilde{f}^\ast_{\rm V}(\mathbf{q};\mathbf{b}_j)}{V_{\rm ea}(\mathbf{b}_i) V_{\rm ea}^\ast(\mathbf{b}_j)} \right)
\right]    \\
\nonumber &= \frac{1}{2}\frac{\lambda^2}{(2\pi)^4} \int d^2 \mathbf{q}\, Q(\mathbf{q}) 
\sum_{i,j=1}^3  {\rm Re}\left[ 
\frac{\tilde{f}_{\rm V}(\mathbf{q};\mathbf{b}_i) \tilde{f}^\ast_{\rm V}(\mathbf{q};\mathbf{b}_j)}{V_{\rm ea}(\mathbf{b}_i) V_{\rm ea}^\ast(\mathbf{b}_j)}  
- 
\frac{\tilde{f}_{\rm V}(\mathbf{q};\mathbf{b}_i) \tilde{f}_{\rm V}(-\mathbf{q};\mathbf{b}_j)}{V_{\rm ea}(\mathbf{b}_i) V_{\rm ea}(\mathbf{b}_j)} 
\right].
\end{align}
This final representation can be readily evaluated numerically. This representation does not require that the baselines close.

\begin{figure}[t]
\centering
\includegraphics[width=0.8\textwidth]{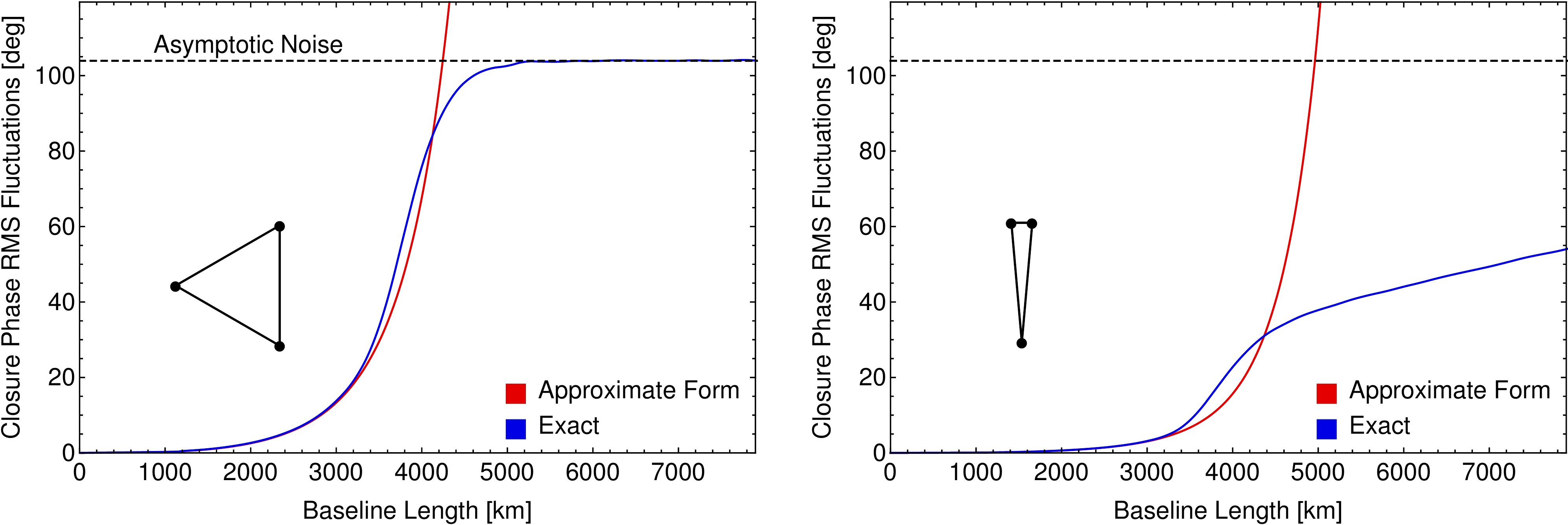}
\caption
{ 
Closure phase fluctuations from refractive scattering as a function of baseline length using our approximate expression (Eq.~\ref{eq::ClosurePhase}) and using the more general Monte Carlo technique (\S\ref{sec::MC}), which is accurate on all baselines. The source and scattering are taken to be isotropic with scales that are comparable to \sgra\ at $\lambda=3.5~{\rm mm}$ \citep{Ortiz_2016}. Specifically, the source is a circular Gaussian with a FWHM of $130~\mu{\rm as}$, and the scattering is isotropic and Kolmogorov with a FWHM of $160~\mu{\rm as}$ (note that the scattering of \sgra\ is anisotropic). The left plot shows results for an equilateral triangle of baselines. The two estimates for closure phase noise agree well almost until the closure phase fluctuations saturate at $60\sqrt{3}$~degrees (corresponding to entirely randomized phase); this transition corresponds to the baselines for which refractive noise is becoming dominant so that the approximation of Eq.~\ref{eq::ClosurePhase} breaks down. For longer baselines, only the Monte Carlo method gives the correct result. The right panel shows the closure phase fluctuations for a long, narrow triangle of baselines with an opening angle of $10^\circ$ (the x-axis baseline length corresponds to the long legs). As for the left panel, the small-noise approximation of Eq.~\ref{eq::ClosurePhase} breaks down when refractive noise on the long legs becomes dominant (near 3500~km), even though the closure phase fluctuations are still small, because the two long legs have highly correlated refractive noise that largely cancels in the closure phase. 
}
\label{fig::SgrA_CPhase2}
\end{figure}

\begin{figure}[t]
\centering
\includegraphics[width=0.48\textwidth]{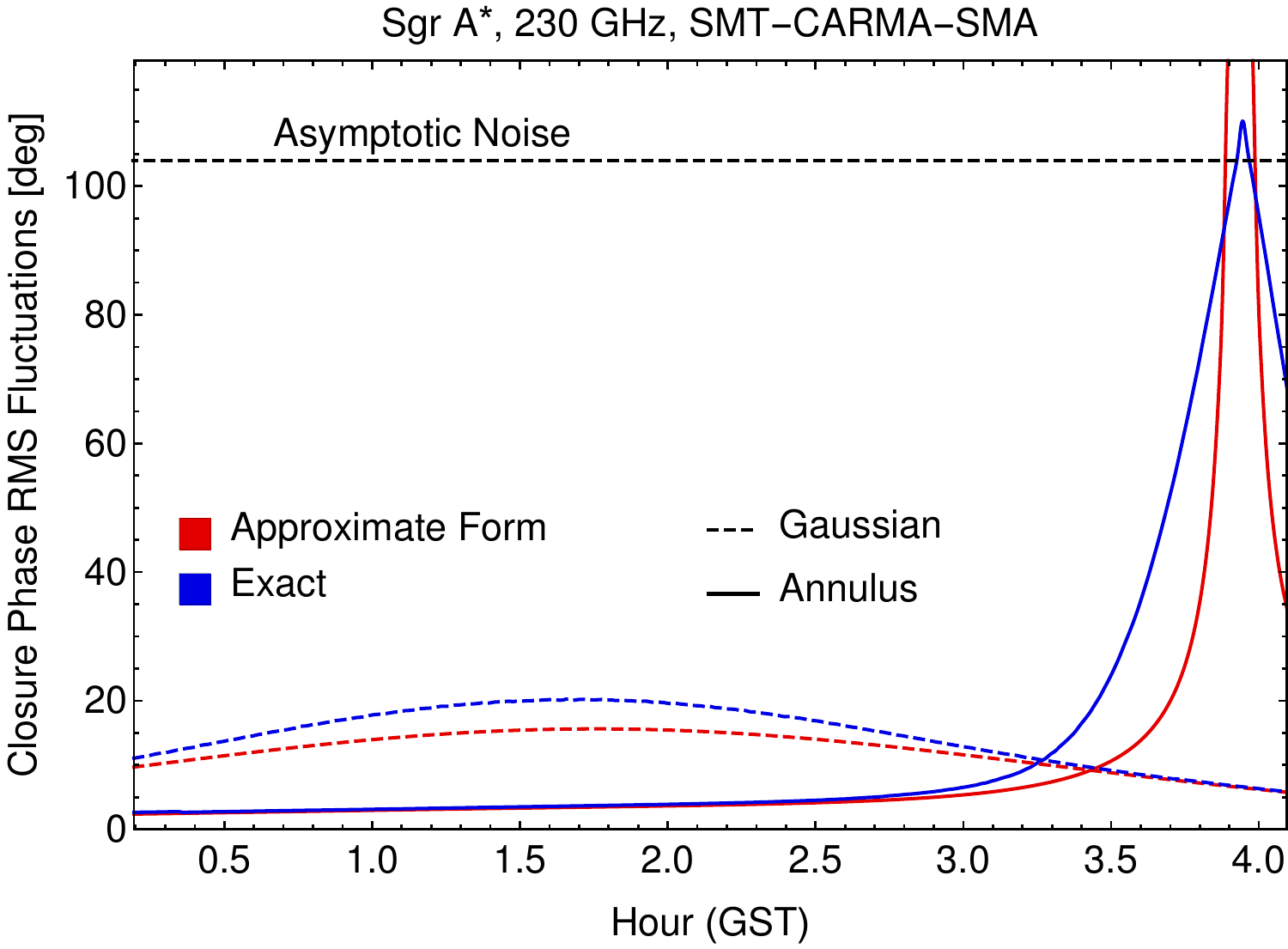}
\caption
{ 
Expected refractive fluctuations of closure phase on the SMT-CARMA-SMA triangle of EHT baselines as a function of Greenwich Sidereal Time (GST) \citep[for recent EHT closure phase measurements on this triangle, see][]{Fish_2016}. As for Figure~\ref{fig::SgrA_CPhase2}, we compare estimates calculated using our small-noise approximation (Eq.~\ref{eq::ClosurePhase}) with the exact values calculated using the Monte Carlo approach outlined in \S\ref{sec::MC}. Two source models are plotted, corresponding to the best-fit Gaussian (FWHM: $52~\mu{\rm as}$) and annulus (inner
diameter: $21~\mu{\rm as}$, outer diameter: $97~\mu{\rm as}$) shown in Figure~S5 of \citet{Johnson_2015}. Note that only the annulus model provides an acceptable fit to the EHT visibility amplitudes reported in \citet{Johnson_2015}. For the annulus, the closure phase RMS fluctuations become large at the end of the track because the SMA-CARMA baseline is approaching a ``null'' in the visibility amplitude. The RMS noise even exceeds the limit of completely random phase because the distribution of closure phase is bimodal, with peaks at $0^\circ$ and $180^\circ$. Because the current visibility amplitudes follow the annulus model, we expect large refractive fluctuations in the closure phase on this triangle (and others that include CARMA-SMA) for GST greater than ${\sim}3.5$.
}
\label{fig::SgrA_CPhase}
\end{figure}

Although Eq.~\ref{eq::ClosurePhase} provides a computationally efficient method to estimate closure phase ``jitter'' from refractive scattering, it requires $|\Delta V_i / V_i| \ll 1$ for \underline{each} of the three participating baselines. Note that there are cases for which the approximation is poor even though the closure phase fluctuations are small. For instance, in a long narrow triangle of baselines, the phase fluctuations on each long baseline may be large individually but will be highly correlated so that the closure phase fluctuations may be small. In these cases, it is advantageous to use the more general approach that we have outlined in \S\ref{sec::MC} to estimate closure phase fluctuations. Although that approach requires a factor of 36 more computations (to evaluate the covariance matrix for the three complex noise terms), it does not require $|\Delta V_i / V_i| \ll 1$. Figures~\ref{fig::SgrA_CPhase2} and \ref{fig::SgrA_CPhase} compare these two approaches to estimate the refractive fluctuations of closure phase for \sgra\ at $\lambda=3.5$ and 1.3~mm, respectively.

Lastly, we caution that refractive fluctuations in closure phase are highly sensitive to the detailed structure of the unscattered image, especially near visibility ``nulls'' of the unscattered image, because they depend on the ratio of refractive noise to the ensemble-average visibility (see Figure~\ref{fig::SgrA_CPhase}). Thus, epoch-to-epoch fluctuations in closure phase can potentially be utilized for model discrimination. In contrast, the RMS fluctuation of visibilities (\S\ref{sec::sigma_ref}) depends on the overall image extent but is not sensitive to the detailed intrinsic structure. Consequently, metrics other than closure phase fluctuations may provide more reliable connections between theory and observations for single-epoch observations, especially when the intrinsic structure is not well-known a priori. Alternatively, if the phase fluctuations are small then one could use measured visibilities as estimates of the ensemble average visibilities $V_{\rm ea}(\mathbf{b}_i)$ in Eq.~\ref{eq::ClosurePhase} to avoid estimates of $\sigma_{\rm CP}$ that are overly model-specific.  


\subsection{The Coherence Timescale for Refractive Noise}

Refractive fluctuations in complex visibilities will change over time from relative motions of the Earth, scattering screen, and source, as discussed in \S\ref{sec::ISM}. The statistical properties of this evolution in time $t$ can be estimated using the frozen-screen approximation: $\phi(\mathbf{r},t+\Delta t) = \phi(\mathbf{r} - \mathbf{V}_\perp \Delta t,t)$. The modification to Eqs.~\ref{eq::BN_Reduction} and \ref{eq::BN_Reduction_Conj} for a pair of phase screens separated by a time $\Delta t$ is straightforward -- the integrands simply obtain extra factors of $e^{-i \mathbf{q} \cdot \mathbf{V}_\perp \Delta t}$. It is then straightforward to estimate the temporal correlation function of refractive metrics \citep[see, e.g.,][]{Blandford_Narayan_1985,Romani_1986}.

Even without detailed computations, we can estimate the relevant coherence timescales from this additional factor; the coherence timescale will be determined by the condition $ |\mathbf{q} \cdot \mathbf{V}_\perp \Delta t| = 1$ for the particular scales $\mathbf{q}$ that dominate the refractive metric of interest. For a point source, refractive metrics such as flux modulation are determined by fluctuations on scales $q \sim 1/r_{\rm R}$, so they evolve on the refractive timescale $t_{\rm R} = r_{\rm R}/V_\perp$. Refractive noise on a baseline of length $\mathbf{b}$ is sensitive to power in the image on scales corresponding to the baseline resolution (see \S\ref{sec::Interferometry}), so they will evolve on a shorter timescale of $\frac{(1+M)^{-1}r_0}{|\mathbf{b}|} t_{\rm R}$. An extended source \emph{increases} the coherence timescales on short baselines by a factor of $\theta_{\rm img}/\theta_{\rm scatt}$ but does not affect the coherence timescale for baselines that resolve the ensemble-average scattered image. Note that the coherence length for displacing a baseline (${\sim} r_{\rm R}$) is much larger than the coherence length for changing a baseline length or orientation (${\sim} r_{\rm img}$; see Figures~\ref{fig::RefractiveNoise_Correlation} and \ref{fig::RefractiveNoise_Correlation_Anisotropic}).

\section{Summary}
\label{sec::Summary}

With rapid increases in angular resolution and sensitivity, effects of interstellar scattering are newly apparent in a variety of VLBI observations \citep[e.g.,][]{Gwinn_2014,Johnson_2016,Ortiz_2016}. For the EHT, imminent additions, notably the Atacama Large Millimeter/submillimeter Array (ALMA) \citep{Fish_2013}, will enable images of \sgra\ at resolution comparable to its event horizon, and understanding how scattering affects these images is essential. For {\it RadioAstron}, the detection of brightness temperatures $T_{\rm b} \gsim 10^{13}~{\rm K}$ at wavelengths from 1.3 to 18~cm \citep{Gomez_2016,Kovalev_2016,Johnson_2016} likewise necessitates a detailed understanding of the role of scattering in these measurements.

The framework that we have developed enables straightforward calculation of refractive scattering effects for these observations. In particular, we have shown how to estimate refractive fluctuations of VLBI observables such as closure phases and amplitudes --- estimates that previously required expensive numerical simulations. Our results accommodate arbitrary source structure and anisotropic scattering. And although we have assumed thin-screen scattering, our results could be generalized to a thick scattering screen or uniform medium, following the approach of \citet{Romani_1986}. 

Our framework also points to new scattering mitigation strategies for these projects. Specifically, because our approach represents the scattered image entirely in terms of the ensemble-average image and its large-scale, refractive perturbations, it cleanly decouples the deterministic ``blurring'' of scattering from the stochastic large-scale distortions. We will explore new mitigation strategies in detail in a subsequent paper.

\acknowledgements{MDJ thanks the National Science Foundation and the Gordon and Betty Moore Foundation (GBMF-3561) for financial support of this work. RN's research was supported in part by NSF grant AST1312651 and NASA grant TCAN NNX14AB47G. MDJ thanks Lindy Blackburn, Katie Bouman, Katherine Rosenfeld, and Andrew Chael for many valuable discussions. We thank Charles Gammie for helpful comments on the manuscript. We thank the referee for many comments that improved the clarity and presentation of these results, and for pointing out connections to literature focused on the weak scattering regime.  
}

\appendix 

\section{The Separation of Diffractive and Refractive Scales}
\label{sec::ScaleSeparation}

We now provide formal justification for the separation of diffractive and refractive scales. Our argument is similar to an argument presented in Appendix~B of \citet{Romani_1986}, and it utilizes results that were given in \citet{Johnson_Gwinn_2015}. We begin with an expression for the snapshot visibility on an interferometric baseline $\mathbf{b}$ that is centered on the position $\mathbf{b}_0$ in the observing plane. Using the Fresnel diffraction integral for the scalar electric field $\psi(\mathbf{b})$ yields \citep[c.f.,][Eq.~3]{Johnson_Gwinn_2015}: 
\begin{align}
\label{eq::Vsnapshot}
V_{\rm ss}(\mathbf{b};\mathbf{b}_0) &\equiv \left \langle \psi^\ast(\mathbf{b}_0-\mathbf{b}/2) \psi(\mathbf{b}_0+\mathbf{b}/2) \right \rangle_{\rm ss}\\
\nonumber &= \frac{1}{4\pi^2 r_{\rm F}^4} \int d^2 \mathbf{x}_1\, d^2 \mathbf{x}_2\, e^{i \frac{1}{2} r_{\rm F}^{-2} \left[ \left(x_1^2 - x_2^2\right) - \frac{\mathbf{b}}{1+M} \cdot \left( \mathbf{x}_1 + \mathbf{x}_2 \right) + 2\frac{\mathbf{b}_0}{1+M}\cdot\left(\mathbf{b} + \mathbf{x}_1 - \mathbf{x}_2 \right) \right]} e^{i \left[ \phi(\mathbf{x}_1) - \phi(\mathbf{x}_2) \right]} \int d^2\mathbf{s}\, e^{i \frac{1}{\lambdabar R} \left( \mathbf{x}_2 - \mathbf{x}_1 \right) \cdot \mathbf{s}} I_{\rm src}(\mathbf{s})\\
\nonumber &= \frac{1}{4\pi^2 r_{\rm F}^4} \int d^2 \mathbf{x}_1\, d^2 \mathbf{x}_2\, e^{i \frac{1}{2} r_{\rm F}^{-2} \left[ \left(x_1^2 - x_2^2\right) - \frac{\mathbf{b}}{1+M} \cdot \left( \mathbf{x}_1 + \mathbf{x}_2 \right) + 2\frac{\mathbf{b}_0}{1+M}\cdot\left(\mathbf{b} + \mathbf{x}_1 - \mathbf{x}_2 \right) \right]} e^{i \left[ \phi(\mathbf{x}_1) - \phi(\mathbf{x}_2) \right]} V_{\rm src}(\left(1+M\right)\left(\mathbf{x}_2 - \mathbf{x}_1 \right)).
\end{align}
The Fresnel scale is $r_{\rm F} \approx \sqrt{ \lambda_{\rm cm} D_{\rm kpc} } \times 2.2 \times 10^5~{\rm km}$ (see \S\ref{sec::ISM}). Because we are primarily interested in resolved sources in the strong scattering regime, we assume that $|\mathbf{b}| \ll r_{\rm F}$ and that the source visibility function imposes a cutoff so that $|\mathbf{x}_2 - \mathbf{x}_1| \ll r_{\rm F}$. Consequently, displacements of $\mathbf{b}_0$ are only significant when they are much greater than the Fresnel scale, so we can simply take $\mathbf{b}_0 = \mathbf{0}$ for the purpose of studying one realization of the snapshot image. The van Cittert-Zernike theorem then relates the snapshot image to the snapshot visibility (c.f., Eq.~\ref{eq::Vavg}):
\begin{align}
\label{eq::Image_Definition}
I_{\rm ss}(\mathbf{x}) &= \frac{1}{\left(\lambda D \right)^2} \int d^2\mathbf{b}\, V_{\rm ss}(\mathbf{b}) e^{2\pi i \left(\frac{\mathbf{b}}{\lambda}\right) \cdot \left(\frac{\mathbf{x}}{D}\right)} = \frac{1}{\left(\lambda D \right)^2} \int d^2\mathbf{b}\, V_{\rm ss}(\mathbf{b}) e^{\frac{i}{r_{\rm F}^2} \frac{\mathbf{b}}{1+M} \cdot \mathbf{x}},
\end{align}
where $\mathbf{x}$ is a transverse coordinate at the distance $D$ of the scattering screen. The integral over $\textbf{b}$ gives a delta function with argument proportional to $\textbf{x} - (\textbf{x}_1 + \textbf{x}_2)/2$. Thus, the only contribution to the snapshot image at a location $\textbf{x}$ is from pairs of points on the screen that are centered on $\textbf{x}$. We can change to variables given by the average and difference of $\textbf{x}_1$ and $\textbf{x}_2$. Using the delta function to integrate over the former leaves a single remaining integral over $\textbf{y} \equiv \textbf{x}_2 - \textbf{x}_1$:
\begin{align}
\label{eq::Image_ss}
I_{\rm ss}(\mathbf{x}) &= \frac{(1+M)^2}{\left(\lambda D \right)^2} \int d^2\mathbf{y}\, 
V_{\rm src}\left( (1+M) \mathbf{y} \right)
e^{i \left[ \phi\left(\mathbf{x}-{\textstyle{\frac{1}{2}}}\mathbf{y}\right) - \phi\left(\mathbf{x}+{\textstyle{\frac{1}{2}}}\mathbf{y}\right)  \right]}
e^{ -\frac{i}{r_{\rm F}^{2}} \mathbf{y}\cdot \mathbf{x} }\!.
\end{align}
Next, we separate the screen phase $\phi(\mathbf{x})$ into two components $\phi(\mathbf{x}) = \phi_{\rm d}(\mathbf{x}) + \phi_{\rm r}(\mathbf{x})$. The diffractive part, $\phi_{\rm d}(\mathbf{x})$, contains the Fourier modes with small-scale variations ($|\mathbf{q}| > 1/r_{\rm F}$), while the refractive part contains the Fourier modes with large-scale variations ($|\mathbf{q}| < 1/r_{\rm F}$). Eq.~\ref{eq::Image_ss} then becomes
\begin{align}
\label{eq::Image_ss2}
I_{\rm ss}(\textbf{x}) &= \frac{(1+M)^2}{\left(\lambda D \right)^2} \int d^2\textbf{y}\, 
V_{\rm src}\left( (1+M) \textbf{y} \right)
e^{i \left[ \phi_{\rm d}\left(\textbf{x}-{\textstyle{\frac{1}{2}}}\textbf{y}\right) - \phi_{\rm d}\left(\textbf{x}+{\textstyle{\frac{1}{2}}}\textbf{y}\right)  \right]}
e^{i \left[ \phi_{\rm r}\left(\textbf{x}-{\textstyle{\frac{1}{2}}}\textbf{y}\right) - \phi_{\rm r}\left(\textbf{x}+{\textstyle{\frac{1}{2}}}\textbf{y}\right)  \right]}
e^{ -\frac{i}{r_{\rm F}^{2}} \textbf{y}\cdot \textbf{x} }\!.
\end{align}
As discussed at the beginning of this section, we assume that the source visibility provides a cutoff $|\mathbf{y}| \ll r_{\rm F}$, so $\mathbf{x}$ ranges over scales $\gg r_{\rm F}$. We can then replace the diffractive exponential by its ensemble average:
\begin{align}
\left \langle e^{ i \left[ \phi_{\rm d}\left(\textbf{x}-{\textstyle{\frac{1}{2}}}\textbf{y}\right) - \phi_{\rm d}\left(\textbf{x}+{\textstyle{\frac{1}{2}}}\textbf{y}\right)  \right] } \right \rangle = e^{-\frac{1}{2} D_\phi(\mathbf{y})}.
\end{align}
This substitution simply takes the snapshot image to the average image and could also be achieved by a partial average in time. Meanwhile, because $|\mathbf{y}|$ is much smaller than the scales of variation of $\phi_{\rm r}$, we can expand the refractive term to leading order: $\phi_{\rm r}(\textbf{x}+\textstyle{\frac{1}{2}}\textbf{y}) - \phi_{\rm r}(\textbf{x}-\textstyle{\frac{1}{2}}\textbf{y}) \approx \textbf{y} \cdot \nabla \phi_{\rm r}(\textbf{x})$. Substituting these diffractive and refractive simplifications into Eq.~\ref{eq::Image_ss2}, we obtain
\begin{align}
\label{eq::Diff_Ref_Split}
I_{\rm avg}(\mathbf{x}) &\approx \frac{(1+M)^2}{\left(\lambda D \right)^2} \int d^2\textbf{y}\, 
V_{\rm src}\left( (1+M) \textbf{y} \right) e^{-\frac{1}{2} D_\phi(\mathbf{y})} 
e^{i \textbf{y} \cdot \nabla \phi_{\rm r}(\textbf{x}) }
e^{ -\frac{i}{r_{\rm F}^{2}} \textbf{y}\cdot \textbf{x} } \\
\nonumber &= \frac{1}{\left(\lambda D \right)^2} \int d^2\textbf{b}\, 
V_{\rm ea}\left(\textbf{b} \right) 
e^{ -\frac{i}{r_{\rm F}^{2}} \frac{\textbf{b}}{1+M}\cdot \left( \textbf{x} + r_{\rm F}^2 \nabla \phi_{\rm r}(\textbf{x}) \right)} \\
\nonumber &= I_{\rm ea}\left( \mathbf{x} + r_{\rm F}^2\nabla\phi_{\rm r}(\mathbf{x}) \right).
\end{align}
The equality on the second line follows from changing variables to $\mathbf{b} = (1+M)\mathbf{y}$ and using the relationship between the visibilities of the unscattered and ensemble-average images (Eq.~\ref{eq::Vea}), and the equality on the third line follows immediately from the van Cittert-Zernicke theorem (see Eq.~\ref{eq::Image_Definition}). Eq.~\ref{eq::Diff_Ref_Split} is equivalent to the result from geometrical optics (Eq.~\ref{eq::RefractiveSteering}). 

\ \\

\bibliography{Substructure_Optics.bib}

\end{document}